\RequirePackage{lineno}
\documentclass[aip,pop,superscriptaddress,reprint]{revtex4-1}

\usepackage[dvipdfmx]{graphicx,color}% Include figure files
\usepackage{dcolumn}% Align table columns on decimal point
\usepackage{bm}% bold math
\begin{document}

\title{
Full particle-in-cell simulation of the interaction between two plasmas 
for laboratory experiments on the generation of magnetized collisionless shocks with high-power lasers 
}

\author{Takayuki Umeda}
\email[Email:]{taka.umeda@nagoya-u.jp}
\affiliation{Institute for Space-Earth Environmental Research, 
Nagoya University, Nagoya 464-8601, JAPAN}

\author{Ryo Yamazaki}
\email[Email:]{ryo@phys.aoyama.ac.jp}
\affiliation{Department of Physics and Mathematics, Aoyama Gakuin University, 
Sagamihara 252-5258, JAPAN}

\author{Yutaka Ohira}
\email[Email:]{y.ohira@eps.s.u-tokyo.ac.jp}
\affiliation{Department of Earth and Planetary Science, The University of Tokyo, 
Tokyo 113-0033, JAPAN}

\author{Natsuki Ishizaka}
\affiliation{Department of Physics and Mathematics, Aoyama Gakuin University, 
Sagamihara 252-5258, JAPAN}

\author{Shin Kakuchi}
\affiliation{Department of Physics and Mathematics, Aoyama Gakuin University, 
Sagamihara 252-5258, JAPAN}

\author{Yasuhiro Kuramitsu}
\affiliation{Graduate School of Engineering, Osaka University, %2-1 Yamadaoka, Suita, 
Osaka 565-0871, JAPAN}

\author{Shuichi Matsukiyo}
\affiliation{Faculty of Engineering Sciences, Kyushu University, %6-1 Kasuga-Koen, 
Kasuga 816-8580, JAPAN}

\author{Itaru Miyata}
\affiliation{Department of Physics and Mathematics, Aoyama Gakuin University, 
Sagamihara 252-5258, JAPAN}

\author{Taichi Morita}
\affiliation{Faculty of Engineering Sciences, Kyushu University, %6-1 Kasuga-Koen, 
Kasuga 816-8580, JAPAN}

\author{Youichi Sakawa}
\affiliation{Institute of Laser Engineering, Osaka University, %2-6 Yamadaoka, Suita, 
Osaka 565-0871, JAPAN}

\author{Takayoshi Sano}
\affiliation{Institute of Laser Engineering, Osaka University, %2-6 Yamadaoka, Suita, 
Osaka 565-0871, JAPAN}

\author{Shuto Sei}
\affiliation{Department of Physics and Mathematics, Aoyama Gakuin University, 
Sagamihara 252-5258, JAPAN}

\author{Shuta J. Tanaka}
\affiliation{Department of Physics and Mathematics, Aoyama Gakuin University, 
Sagamihara 252-5258, JAPAN}

\author{Hirohumi Toda}
\affiliation{Department of Physics and Mathematics, Aoyama Gakuin University, 
Sagamihara 252-5258, JAPAN}

\author{Sara Tomita}
\affiliation{Department of Physics and Mathematics, Aoyama Gakuin University, 
Sagamihara 252-5258, JAPAN}

\newcommand{\Vec}[1]{\mbox{\boldmath $#1$}}

\begin{abstract}
A preliminary numerical experiment is conducted 
for laboratory experiments on the generation of magnetized collisionless shocks 
with high-power lasers by using one-dimensional particle-in-cell simulation. 
The present study deals with the interaction between 
a moving Aluminum plasma and a Nitrogen plasma at rest. 
In the numerical experiment, 
the Nitrogen plasma is unmagnetized or magnetized by a weak external magnetic field. 
Since the previous study suggested the generation of spontaneous magnetic field 
in the piston (Aluminum) plasma due to the Biermann battery,
the effect of the magnetic field is of interest. 
Sharp jumps of electron density and magnetic field are observed 
around the interface between the two plasmas 
as long as one of the two plasmas is magnetized, 
which indicates the formation of tangential electron-magneto-hydro-dynamic discontinuity. 
When the Aluminum plasma is magnetized, 
strong compression of both density and magnetic field takes place in the pure Aluminum plasma 
during the gyration of Nitrogen ions in the Aluminum plasma region. 
The formation of a shock downstream is indicated from the shock jump condition. 
The result suggests that 
the spontaneous magnetic field in the piston (Aluminum) plasma 
plays an essential role in the formation of a perpendicular collisionless shock. 
\end{abstract}

%\pacs{
%52.35.Tc; %Shock waves and discontinuities
%52.65.-y; %Plasma simulation
%52.65.Rr; %Particle-in-cell method
%}

\maketitle

%%%%%%%%%%%%%%%%%%%%%%%%%%%%%%%%%%%%%%%%%%%%%%%%%%%%%%%%%%%%%%%%%%%%%%%

%\linenumbers

\section{Introduction}

Collisionless shocks play important roles in the generation of high-energy particles 
in various situations, 
which is one of the most important outstanding issues in plasma physics.\cite{Balogh_2013,Burgess_2015} 
Recently, laboratory experiments using high-power lasers are conducted 
on the generation of collisionless shocks propagating into 
unmagnetized\cite{Kuramitsu_2010,Morita_2010,Yuan_2017,Ross_2017}
and magnetized\cite{Schaeffer_2014,Niemann_2014,Schaeffer_2017a,Schaeffer_2017b,Shoji_2016} plasmas. 
In particular, the laboratory experiments of magnetized collisionless shocks are of great interest
since the most astrophysical and solar-terrestrial plasmas hosting the collisionless shocks are magnetized.

There are mainly two ways to excite collisionless shocks in laboratory plasma experiments 
using high-energy lasers, 
in which two plasmas collide with each other.
One is to have counter-streaming plasmas, both of which move in the laboratory frame.
They arise from double-plane targets irradiated by lasers.\cite[e.g.,][]{Schaeffer_2017a,Schaeffer_2017b}
In the other way,\cite{Shoji_2016} 
an ambient plasma at rest %is set which 
is pushed by a flowing plasma originated in laser ablation.
To make the ambient plasma, the neutral gas is fulfilled around the target before the shot, 
and it is photoionized by photons generated in the laser ablation process.
The ambient plasma is magnetized if the external magnetic field is imposed before it is ionized. 
In this method, one can easily control the field strength, accordingly 
the Alfv\'{e}n Mach number and the plasma beta 
(i.e., the ratio of the plasma pressure to the magnetic pressure) of the ambient plasma. 
In contrast, some authors have proposed a complementary way toward
the collisionless shock formation using ultra-high-intensity lasers 
to drive a fast quasi-neutral flow in a denser plasma.\cite{Fiuza_2012,Ruyer_2015,Grassi_2017}

It has been believed that in the ablation plasma, spontaneous magnetic fields are produced
by laser-plasma interactions {due to the 
so-called \textit{Biermann battery} process.\cite{Biermann_1950}}
The Biermann battery works when the cross product between the gradients of
the electron density and the electron temperature is non-vanishing near 
the targets.\cite{Stamper_1991,Gregori_2012,Kugland_2013}
The resultant magnetic field is toroidal with respect to the direction of the plasma flow.
Then, the magnetic field convects with the ablation plasma outwards.\cite{Kugland_2013,Ryutov_2013}
At least just after the shot at which strong density and temperature gradients exist,
the magnetic pressure can be comparable to or even larger than the kinetic and thermal pressure
of the plasma.
However, at present, the role of the Biermann magnetic field 
in the excitation of the collisionless shocks is poorly understood.

%%%%%%%%%%%%%%%%%%%%%%%%%%%%%%%%%%%%%%%%%%

\begin{table*}[b]
\caption{
Simulation parameters for the present numerical experiment. 
A reduced value in the numerical experiment is shown 
at the right-hand side when the physical quantity is not the real one. 
The cyclotron frequency, the thermal gyro radius, the Alfv\'{e}n velocity, and the plasma beta 
are for runs with non-zero magnetic field ($B_0\ne0$). 
}
\renewcommand{\arraystretch}{0.75}
\begin{tabular}{c||c|c}
Quantity & Aluminum plasma & Nitrogen plasma \\ \hline \hline
Drift velocity $V_d$ [km/s] & 500 & 0 \\
Magnetic field $B_0$ [T] && \\
Run 1 & 10.0 & 0.5 \\
Run 2 & 10.0 & 0.0 \\
Run 3 & 0 & 0.5 \\
Run 4 & 0 & 0.0 \\
\hline
Electrons & \ \ \ \ \ \ \ \ \ \ \ \ \ \ \ \ \ \ 2,250/cell & \ \ \ \ \ \ \ \ \ \ \ \ \ \ \ \ \ \ 90/cell \\
Density $N_e$ [${\rm cm}^{-3}$] & $3.75\times10^{19}$ & $1.5\times10^{18} $ \\
Plasma frequency $f_{pe}$ [Hz] &  $5.51\times10^{13}$ / $5.51\times10^{12}$ & 
$1.1\times10^{13}$ / $1.1\times10^{12}$ \\ 
Temperature $T_e$ [eV] & 10 & 30 \\
Thermal velocity $V_{te}$ [km/s] & 1,330 & 2,300 \\
Debye length $\lambda_{De}$ [m] & $3.71\times10^{-9}$ / $3.71\times10^{-8}$ &
$3.32\times10^{-8}$ / $3.32\times10^{-7}$ \\
Inertial length $d_e$ [m] &  $8.39\times10^{-7}$ &  $4.33\times10^{-6}$ \\
Cyclotron frequency $f_{ce}$ [Hz] & $2.8\times10^{11}$ & $1.4\times10^{10}$ \\ 
Thermal gyro radius $r_e$ [m] & $7.55\times10^{-7}$ &  $2.62\times10^{-5}$ \\
Plasma beta $\beta_e$ & 1.62 & 72.99 \\
\hline
Ions & \ \ \ \ \ \ \ \ \ \ \ \ \ \ \ \ \ \ 250/cell & \ \ \ \ \ \ \ \ \ \ \ \ \ \ \ \ \ \ 30/cell \\
Charge number $Z$ & 9 & 3 \\
Mass ratio $m_i/m_e$ & 49572 & 25704 \\
Density $N_i$ [${\rm cm}^{-3}$] & $4.17\times10^{18}$ & $5.0\times10^{17}$ \\
Plasma frequency $f_{pi}$ [Hz] & $7.43\times10^{11}$ / $7.43\times10^{10}$ & 
$1.19\times10^{11}$ / $1.19\times10^{10}$ \\ 
Temperature $T_i$ [eV] &10 & 30  \\
Thermal velocity $V_{ti}$ [km/s] & 5.97 & 14.3 \\
Debye length $\lambda_{Di}$ [m] & $3.71\times10^{-9}$ / $3.71\times10^{-8}$ &
$3.32\times10^{-8}$ / $3.32\times10^{-7}$ \\
Inertial length $d_i$ [m] & $6.23\times10^{-5}$ & $4.01\times10^{-4}$ \\
Cyclotron frequency $f_{ci}$ [Hz] & $5.08\times10^{7}$ & $1.63\times10^{6}$ \\ 
Thermal gyro radius $r_i$ [m] & $1.87\times10^{-5}$ & $1.4\times10^{-3}$ \\
Alfv\'{e}n velocity $V_A$ [km/s] & 19.99 & 4.11 \\
Plasma beta $\beta_i$ & 0.18 & 24.33 \\
\hline
Grid spacing $\Delta x$ [m]& \multicolumn{2}{c}{$8.3 \times 10^{-8}$} \\
Time step $\Delta t$ [sec]& \multicolumn{2}{c}{$2.6 \times 10^{-15}$} \\ 
Number of grids $N_x$ & \multicolumn{2}{c}{120,000} \\
Number of steps $N_t$ & \multicolumn{2}{c}{6,000,000} \\
Speed of light $c$ [km/s]& \multicolumn{2}{c}{300,000 / 30,000} \\
& \multicolumn{2}{c}{(laboratory) / (numerical)} \\
\end{tabular}
\end{table*}

%%%%%%%%%%%%%%%%%%%%%%%%%%%%%%%%%%%%%%%%%%

In the present study, 
we perform one-dimensional (1D) full particle-in-cell (PIC) simulations 
of the interaction between the ablation (piston) plasma and the ambient plasma at rest,  
as a preliminary numerical experiment for laboratory experiment 
on the generation of magnetized collisionless shocks with high-power lasers such as
Gekko-XII at  the Institute of laser engineering in Osaka University.  
%
%Due to a busy schedule of the laser facilities, 
%laboratory experiments cannot be conducted so frequently. 
Since the number of shots is limited in laboratory experiments,
our preliminary numerical experiments play {an important role} 
in the prediction of results of laboratory experiments. 
The present study aims to study the effect of the spontaneous magnetic field 
in the piston plasma {due to the Biermann battery} 
on the generation of collisionless shocks.

\section{1D PIC Simulation}

\subsection{Typical Parameters}

First, we briefly describe our setup and accompanying typical parameters
optimized for laboratory experiments using Gekko-XII HIPER lasers.
Details will be published elsewhere. 
The planar Aluminum target is irradiated by HIPER lasers with energy of $\sim$~kJ in total
and the pulse width of $\sim$~nsec, producing the Aluminum plasma with bulk
velocity of $\sim10^2$~km/s.
A vacuum chamber is filled with Nitrogen gas with 5~Torr before the shot, and it is ionized
by photons arising in the laser-target interaction.
Just after the shot, the Aluminum plasma is very hot and dense with electron density
$N_e\sim10^{21}$~${\rm cm}^{-3}$
and temperature $T_e\sim10^3$~eV near the target.
As it expands, it adiabatically cools down and has typically $N_e\sim10^{19}$~${\rm cm}^{-3}$
and $T_e\sim10$~eV when the shock is formed. 
The Nitrogen plasma is cold with temperature $T_e\sim$~eV at the beginning.
However, it is preheated by HIPER lasers to about several tens of eV 
since we consider interaction between Aluminum 
and Nitrogen plasmas in the ablation side of the target.
%\textcolor{red}{Schematic views of our preliminary laboratory experiments 
%are shown in Ref.\cite{Shoji_2016}. }

Unfortunately, the strength of the Biermann spontaneous magnetic field 
associated with the Aluminum plasma 
is unknown since magnetic fields were not measured in our preliminary laboratory experiments. 
Here, we roughly estimate the strength of the spontaneous magnetic field as follows. 
The evolution of the magnetic field is described by the following equation {in SI units 
(so that the Boltzmann constant is omitted)}
\begin{equation}
\frac{\partial \Vec{B}}{\partial t} \approx\frac{1}{eN_e} \left(\nabla T_e \times \nabla N_e \right),
\end{equation}
%where $k_B$ and $e$ are the Boltzmann's constant and the electron charge, respectively. 
which is derived from the rotation of the {electron pressure gradient term 
of electric fields, i.e., $-\nabla \times \{\nabla P_e/(eN_e)\}$} 
in the magnetic induction equation. 
Here, we {have neglected} the convection term, $\nabla\times(\Vec{U}_e \times \Vec{B})$, 
at the very beginning of the field generation.
If we approximate
$\partial/\partial t\approx V_d/\phi$ and $\nabla\approx1/\phi$, where
$V_d$ and $\phi$ are the drift velocity of the piston Aluminum plasma and 
the focal spot size of HIPER lasers, respectively, 
then we have
\begin{eqnarray}
B&\approx&\frac{T_e}{eV_d\phi} \nonumber\\
&\approx& 10~{\rm T}
\left(\frac{T_e}{10^3~{\rm eV}}\right)
\left(\frac{V_d}{10^2~{\rm km/s}}\right)^{-1}
\left(\frac{\phi}{1~{\rm mm}}\right)^{-1}.
\end{eqnarray}
Inserting physical parameters near the target into Eq.(2), 
we obtain a typical magnitude of the Biermann spontaneous magnetic field 
as $\sim 10$ T, which is consistent with other laboratory experiment\cite{Fiksel_2014}
and numerical experiment.\cite{Fox_2018}

On the other hand, the external magnetic field imposed on the ambient Nitrogen plasma 
has arbitrary strength. 
However, its typical value is $B\sim1$~T.
If $B\ll1$~T, then the ion is not magnetized.
If $B\gg1$~T, then the magnetized shock propagating into the Nitrogen plasma
has a small Alfv\'{e}n Mach number.
{In the present study, we assume an unmagnetized ambient Nitrogen plasma 
as shown in Table.1.}

%%%%%%%%%%%%%%%%%%%%%%%%%%%%%%%%%%%%%%%%%%%%%%%%

\subsection{Simulation Setup}

We use a 1D relativistic full PIC code developed by ourselves 
that was used for simulations of collsionless shocks.\cite{Umeda_2006}  
{The Sokolov interpolation \cite{Sokolov_2013} is implemented into 
the second-order charge conservation scheme\cite{Umeda_2003}} 
to reduce numerical noises.
The code has stable open boundary conditions 
which allows us to perform simulations with a long time of $t \gg 10^5/\omega_{pe}$. 

In the present numerical experiment, 
the simulation domain is taken along the $x$ axis. 
At the initial state,  
the simulation domain is filled with collisionless Nitrogen plasma at rest as an ambient plasma. 
The open boundary conditions are imposed at both of boundaries, 
where electromagnetic waves and plasma particles escape freely. 
In addition to the open boundary condition, 
collisionless Aluminum plasma as a piston plasma with a drift velocity $V_d$ 
is continuously injected from at the left boundary ($x=0$) into the Nitrogen plasma
when the numerical experiment is started. 
All of the plasma particles have a (shifted) Maxwellian velocity distribution 
with an isotropic temperature at the initial state. 

The typical parameters of the Aluminum plasma and the Nitrogen plasma 
{near a measurement point of our preliminary laboratory experiments 
($\sim$ cm away from the target)} are listed in Table 1.  
{In the present numerical experiment, 
we have tried to use these parameters as many as possible, 
including the real ion-to-electron mass ratios. }
However, some of them are reduced with respect to the real ones 
to save computational cost. 
As an example, 
the vacuum permittivity $\epsilon_0$ used in the present numerical experiment is 
100 times larger than the real one. 
This means that the speed of light and the plasma frequency are 
reduced to one-tenth of the real ones. 
However, the Alfv\'{e}n velocity, plasma beta and the inertial length are 
set to be the same as real ones, which are important parameters for discussion of the shock dynamics. 
{In Table 1, the real values in the laboratory experiments are shown 
at the left-hand side and reduced values are shown at the right-hand side when the physical quantity 
in the numerical experiment is reduced. }
These parameters are renormalized to the electron plasma frequency and 
the electron thermal velocity in the Nitrogen plasma in the full PIC simulation. 
{The number of particles per cell for each species is also shown in Table 1}.

The ambient magnetic field $B_0$ (in the Nitrogen plasma) is imposed in the $y$ direction 
with a magnitude of 0.5 T, 
in contrast to the previous studies in which the magnitude of the ambient magnetic field 
was 4 T.\cite{Schaeffer_2017a,Schaeffer_2017b}
Since the magnitude of the ambient magnetic field is weak, 
the ambient (Nitrogen) plasma is in the high beta regime in the present study. 
As a spontaneous magnetic field due to the Biermann battery, 
we assume that the magnetic field in the piston (Aluminum) plasma 
is directed in the $y$ direction with a magnitude of 10 T. 
To magnetize the Aluminum plasma, a motional electric field is 
also imposed in the $z$ direction with a magnitude of $E_z=-V_dB_0$ at the left boundary ($x=0$).

We perform four different runs. 
In Run 1, both of the Aluminum plasma and the Nitrogen plasma are magnetized. 
In Run 2, the Aluminum plasma is magnetized while the Nitrogen plasma is unmagnetized. 
Note that it is easy to change the magnitude of the ambient magnetic field (in the Nitrogen plasma) 
in laboratory experiments. 
In Run 3, the Nitrogen plasma is magnetized while 
the magnetic field in the Aluminum plasma is set to be zero 
to see the effect of the spontaneous magnetic field in the piston plasma. 
In Run 4, both of the Aluminum plasma and the Nitrogen plasma are unmagnetized. 
The direct comparison among these runs could show the influence of 
the spontaneous magnetic field in the Aluminum plasma 
{due to the Biermann battery} 
to the generation of collsionless shocks.

\section{Simulation Result}

\begin{figure}[p]%[t]
\center
\includegraphics[width=1.0\textwidth,bb=0 0 150 150]{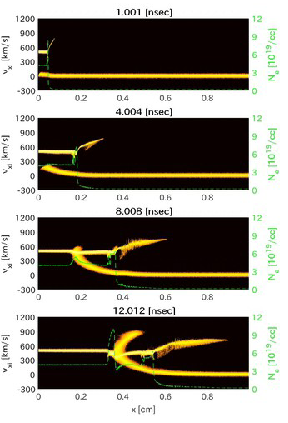}
\caption{
Temporal development of the interaction between 
the Aluminum plasma and the Nitrogen plasma for Run 1. 
The $x-v_x$ phase-space plots of ions at different times 
together with the spatial profile of the electron density. 
}
\end{figure}

Figure 1 shows the temporal development of the interaction between 
the Aluminum plasma and the Nitrogen plasma for Run 1. 
The \textit{ion charge density} in the $x-v_x$ phase space is plotted at different times. 
The corresponding spatial profile of the electron density is also superimposed. 
Since the absolute value of the electron charge density is almost equal to the 
total ion charge density, i.e., the sum of charge densities of Aluminum and Nitrogen ions, 
the quasi charge neutrality is almost satisfied at all the time. 

At the leading edge of the Aluminum plasma, there exists strong charge separation 
because the Aluminum ions can penetrate the Nitrogen plasma region 
while electrons cannot compensate 
the Aluminum ion charge owing to the small gyro radius. 
A diamagnetic current is also generated 
around the interface between the two plasmas due to a large magnetic shear. 
The charge separation and electric current excite electromagnetic fluctuations, 
which results in the ponderomotive force to scatter (accelerate) Aluminum ions  
at the leading edge toward the Nitrogen plasma region. 
The ponderomotive force also reflect some of Aluminum ions around the interface 
toward the Aluminum plasma region. 
Note that the acceleration of electrons due to the ponderomotive force is not seen 
since accelerated electrons soon diffuse in the velocity space through gyration. 

As the Aluminum plasma penetrates into the Nitrogen plasma region, 
an instability is generated at the leading edge of the Aluminum plasma 
(see second and third panels at $t=4.004$ and $8.008$ nsec, respectively). 
We found a wave mode is excited at the local electron cyclotron frequency $\omega_{ce,local}$ 
from the Fourier analysis. 
The wavenumber satisfies the resonance condition $k_xV_{dAl} \approx \omega_{ce,local}$, 
suggesting that the electron cyclotron drift instability is generated due to 
the drift of Aluminum ions across the ambient magnetic field. 
{For more detail, see Appendix A.}

As the time elapses, the Nitrogen ions feel the motional electric field 
in the Aluminum plasma region and gyrate in the $x-v_x$ phase space. 
During the gyration, strong compression of Aluminum ions takes place 
in a region outside the gyrating Nitrogen ions 
(at $x \approx 0.35$ cm at $t = 12.012$ nsec).

\begin{figure}[t]
\center
\includegraphics[width=1.0\textwidth,bb=0 0 150 150]{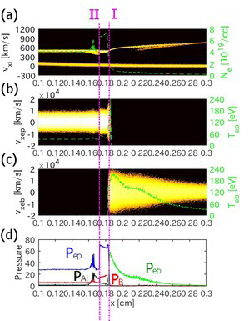}
\caption{
(a) Ion $x-v_x$ phase space together with the spatial profile of the electron density 
around the interface between the Aluminum plasma and the Nitrogen plasma  
at $t = 4.004$ nsec for Run 1. 
(b) Electron $x-v_x$ phase space together with the spatial profile of the electron temperature 
of the Aluminum (piston) plasma.  
(c) Electron $x-v_x$ phase space together with the spatial profile of the electron temperature 
of the Nitrogen (background) plasma.  
(d) Spatial profiles of the pressures of the piston electrons $P_{ep}$, the background electrons $P_{eb}$, 
the Aluminum ions $P_{Al}$, and the magnetic pressure $P_B$. 
The pressure is normalized to the initial electron pressure in the Nitrogen (background) plasma region. 
}
\end{figure}

Panel (a) of Fig.2 shows an expansion of the ion $x-v_x$ phase space 
together with the  spatial profile of the electron density 
around the interface between the two plasmas at $t = 4.004$ nsec for Run 1.  
Panels (b) and (c) show the corresponding 
electron $x-v_x$ phase space together with the  spatial profile 
of the electron temperature of the Aluminum (piston) and 
the Nitrogen (background) plasmas, respectively. 
Panel (d) shows the  spatial profiles of the pressures of the piston electrons $P_{ep}$, 
the background electrons $P_{eb}$, 
the Aluminum ions $P_{Al}$, and the magnetic pressure $P_B=|B_y|^2/(2\mu_0)$ 
normalized to the initial electron pressure of the Nitrogen (background) plasma region. 
Note that the plasma pressure is given as $P=(P_{x}+P_{z})/2$. 
The pressure of the Nitrogen ions is small and is not shown here. 
The magnetic field $B_y$ component is almost proportional 
to the total (the sum of piston and background) electron density 
and is not shown here. 

There are two discontinuous structures (labeled as ``I'' and ``II'') 
around the interface between the two plasmas. 
We found that the piston electrons are clearly separated 
from the background electrons around the discontinuity ``I'' 
as seen in panels (b) and (c) of Fig.2. 
The electron density, the electron temperature, and the electron pressure 
continuously increase from the right to the interface between the two plasmas. 
The density of piston electrons increases from the right to the left 
but the temperature of piston electrons decreases at the discontinuity ``I''. 
The pressure of piston electrons decreases slightly 
(see ``$P_{ep}$'' in Panel (d)) but the magnetic pressure increases at the discontinuity ``I''. 
It is suggested that the electron pressure gradient force balances the $\Vec{J}_e\times \Vec{B}$ force. 
Since the bulk velocities of piston and background electrons are almost the same 
across the discontinuity ``I'', 
the discontinuity ``I'' can be regarded as a tangential 
electron-magneto-hydro-dynamic (EMHD) discontinuity. 

Panel (c) of Fig.2 shows that the background electrons are heated 
due to the penetration of the piston (Aluminum) ions. 
The mechanism of the electron heating is considered to be the following 
EMHD process, since the timescale of the 
diffusion of electrons in the velocity space due to the gyration is fast. 
The background electrons feel the motional electric field of piston ions, 
which result in a finite bulk velocity of the background electrons ($U_{xeb} > 0$).  
Then, the pressure of the background electrons increases from the right to the left, as seen in panel (d), 
by the compression of the background electrons, i.e., $\partial U_{xeb}/\partial x \ne 0$. 
The sum of magnetic pressure and the thermal pressure of electrons (i.e., $P_B+2P_{ep}$) is 
kept almost constant across the discontinuity ``I''. 
Since the electron density of the piston plasma is higher than 
that of the background plasma, there exists a discontinuity of the electron temperature 
%background electrons are heated apparently 
around the interface between the two plasmas. 

At the discontinuity ``II'', the total electron density decreases from the right to the left 
but the pressure of Aluminum ions increases. 
The sum of the pressure of Aluminum ions and the magnetic pressure 
is kept almost constant across the discontinuity ``II''.

\begin{figure}[t]
\center
\includegraphics[width=1.0\textwidth,bb=0 0 150 150]{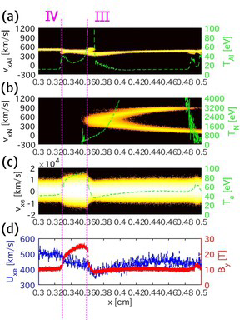}
\caption{
(a) Ion $x-v_x$ phase space together with the ion temperature of the Aluminum plasma 
around the gyrating Nitrogen ions at $t = 12.012$ nsec for Run 1, 
and (b) the corresponding ion $x-v_x$ phase space together 
with the spatial profile of the ion temperature for the Nitrogen plasma. 
(c) Electron $x-v_x$ phase space together with the spatial profile of the electron temperature.  
(d) Spatial profiles of the electron bulk velocity $U_{xe}$ and the magnetic field $B_y$.  
}
\end{figure}

Panels (a) and (b) of Fig.3 show an expansion of the ion $x-v_x$ phase space 
together with the  spatial profile of the ion temperature of the Aluminum and Nitrogen plasmas, respectively, 
around the gyrating Nitrogen ions at $t = 12.012$ nsec for Run 1.  
Panel (c) shows the corresponding 
electron $x-v_x$ phase space together with the  spatial profile of the electron temperature. 
Panel (d) shows the spatial profiles of the electron bulk velocity $U_{xe}$ and the magnetic field $B_y$.  
Note that electrons in this region consist of piston electrons. 

There are two discontinuous structures (labeled as ``III'' and ``IV'') 
around the gyrating Nitrogen ions. 
At the discontinuity ``III'', the electron density, the electron temperature, and 
the magnetic field increase from the right to the left. 
The electron bulk velocity changes from $\approx 400$ km/s to $\approx 450$ km/s. 
At the discontinuity ``IV'', the electron density, the electron temperature, and 
the magnetic field decrease from the right to the left.  
The electron bulk velocity changes from $\approx 450$ km/s to $\approx 500$ km/s. 
The density and the magnetic field between the discontinuities ``III'' and ``IV'' 
are $\approx 2.4$ times as large as those in the unperturbed Aluminum plasma.

Let us consider the MHD shock jump conditions for perpendicular shocks, 
\begin{eqnarray}
&&B_{y1}U_1 = B_{y2}{U_2}, \\
&&N_1U_1 = N_2 U_2, \\
&&(m_e+m_i)N_1U_1^2+P_{1}+\frac{B_{y1}^2}{2\mu_0} \\ \nonumber
&& \ = 
(m_e+m_i)N_2U_2^2+P_{2}+\frac{B_{y2}^2}{2\mu_0}, \\
&&\left\{\frac{(m_e+m_i)N_1U_1^2}{2}+2P_1+\frac{B_{y1}^2}{\mu_0}\right\}U_1 \\ \nonumber
&&=
\left\{\frac{(m_e+m_i)N_2U_2^2}{2}+2P_2+\frac{B_{y2}^2}{\mu_0}\right\}U_2,
\end{eqnarray}
where, the subscript ``1'' and ``2'' denotes the upstream and the downstream of a discontinuity, respectively. 
Here, the ion and electron densities are assumed to be almost equal, $N\equiv N_i \approx N_e$, 
and the plasma pressure is given as the sum of ion and electron pressures, $P\equiv P_i+P_e$. 
The bulk velocity $U$ is defined in rest frame of a discontinuity.

Suppose that the velocity of the discontinuities ``III'' and ``IV'' is 
$\approx 485.7$ km/s and $\approx 414.3$ km/s, respectively. 
Then, Eqs.(3) and (4) are satisfied across the discontinuities ``III'' and ``IV'' with 
{the upstream bulk velocity of $U_1\approx 85.7$ km/s, 
the downstream bulk velocity of $U_2\approx 35.7$ km/s 
and the density $N_2/N_1\approx 2.4$} 
in the rest frame of the discontinuities. 
{From the initial conditions, we obtain $(m_e+m_i)N_1U_1^2/P_{i1} \approx 206$, 
$P_{e1}/P_{i1} = 9$ and $B_{y1}^2/(2\mu_0P_{i1}) \approx 5.6$. 
The pressures of the electrons and Aluminum ions in 
the high density region (downstream) 
are $P_{e2}/P_{e1} \approx 11$ and $P_{Al2}/P_{Al1} \approx 8$
times as large as those of the unperturbed Aluminum plasma (upstream), respectively. 
Then, we obtain $(m_e+m_i)N_2U_2^2/P_{i1} \approx 85.8$, $P_{2}/P_{i1} \approx 92$ 
and $B_{y2}^2/(2\mu_0P_{i1}) \approx 32$. 
The momentum conservation law and the energy conservation law 
for the Aluminum plasma in Eqs.(5) and (6), respectively, are almost satisfied.  }
Hence, the shock jump conditions are almost satisfied across the discontinuities ``III'' and ``IV''. 

This result suggests that 
these discontinuous structures correspond to collisionless perpendicular shocks. 
The high-density region between the discontinuities ``III'' and ``IV'' corresponds to the shock downstream. 
The typical Alfv\`{e}n Mach number of the shock is $M_A \approx 4.2$. 
The shock downstream is located at a distance of one ion gyro radius of Nitrogen ions in the Aluminum plasma from the interface between the Nitrogen and Aluminum plasmas. 
The shock is formed on the timescale of a quarter ion gyro period of Nitrogen ions in the Aluminum plasma.

\begin{figure*}[p]
\center
\includegraphics[width=0.9\textwidth,bb=0 0 200 200]{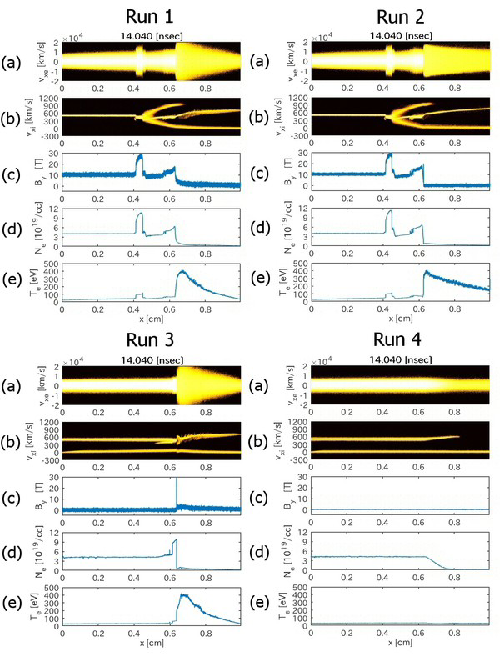}
\caption{
Snapshot of four Runs at $t = 14.040$ nsec. 
(a) The $x-v_x$ phase-space plots of electrons. 
(b) The $x-v_x$ phase-space plots of ions. 
(c) The spatial profile of the magnetic field $B_y$ component. 
(d) The spatial profile of the electron density $N_e$. 
(e) The spatial profile of the electron temperature $T_e$. 
}
\end{figure*}

Figure 4 shows a snapshot of four Runs at $t = 14.040$ nsec. 
Panel (a) shows the $x-v_x$ phase-space plots of electrons. 
Panel (b) shows the $x-v_x$ phase-space plots of ions. 
Panel (c) shows the spatial profile of the magnetic field $B_y$ component. 
Panel (d) shows the spatial profile of the electron density $N_e$. 
Panel (e) shows the spatial profile of the electron temperature $T_e$.

In Run 2 where the ambient magnetic field in the Nitrogen plasma is absent, 
the electron cyclotron drift instability is not generated 
at the leading edge of the Aluminum plasma. 
However, electron heating takes place due to the penetration of the Aluminum ions  
(for $x > 0.626$ cm), suggesting that the generation of the instability
is not a necessary condition for the electron heating. 
The electron temperature at the leading edge of the Aluminum plasma in Run 2 
is higher than that in Run 1, since electrons in this region is unmagnetized and 
they escape from the interface. 
The tangential discontinuity and the shock downstream are formed at $x \approx 0.55$ cm
and $x \approx 0.41$ -- $0.45$ cm, respectively. 
The comparison between Runs 1 and 2 suggests that 
the existence of the ambient magnetic field (in the Nitrogen plasma) is not a 
necessary condition for the formation of shocks in the Aluminum plasma. 

In Run 3 where the magnetic field in the Aluminum plasma is absent, 
similar electron heating also takes place due to the penetration of the Aluminum ions  
(for $x > 0.637$ cm). 
The electron cyclotron drift instability is generated as in Run 1, 
and the spatial profile of the electron temperature is almost the same as 
that in Run 1. 
A strong fluctuation of the magnetic field is excited at $x \approx 0.637$ cm, 
and a part of Aluminum ions are reflected by the ponderomotive force. 
Since the gyration of Nitrogen ions is absent, the shock downstream is not formed in Run 3.  

The Aluminum plasma pass through the 
Nitrogen plasma, and no interaction between them is seen in Run 4. 
The acceleration of Aluminum ions purely by the electrostatic field 
due to charge separation is seen in panel (b), which is weaker than the acceleration 
in the other runs. 
Also, electron heating due to the penetration of the Aluminum ions 
does not take place. 
%These results suggest that 
%the magnetic pressure gradient and 
%the ponderomotive force of electromagnetic fields 
%play important roles in the interaction between the two plasmas. 

{Finally, it should be noted that we also performed several runs 
with a large ambient magnetic field (e.g., 3 T and 5 T) in the Nitrogen plasma region. 
In these runs, a collisionless shock is formed by the gyration of Aluminum ions in the 
Nitrogen plasma region, which is consistent with the previous study.\cite{Schaeffer_2017a,Schaeffer_2017b}
The influence of magnetic field in the Aluminum plasma due to the Biermann battery 
on the formation of discontinuities and shocks is small. 
Hence, these runs with a large ambient magnetic field are out of purpose of the present study and 
are not shown here. 
}

\section{Summary} 

The interaction between the piston Aluminum plasma and the ambient Nitrogen plasma  
was studied by means of a 1D full PIC simulations 
as a preliminary numerical experiment for laboratory experiment 
on the generation of magnetized collisionless shocks with high-power lasers. 
Preliminary numerical experiments are important 
in the prediction of results of laboratory experiments.

Four different runs were performed with the combination of 
{two magnetized and/or unmagnetized plasmas}. 
It is shown that the magnetic field plays an important role 
in the formation of the tangential EMHD discontinuity 
around the interface between the two plasmas 
in the present study with a weak ambient magnetic field. 
The tangential EMHD discontinuity is formed on the timescale of the electron gyro period.
This result is different from the previous result\cite{Schaeffer_2017a,Schaeffer_2017b}
in which a strong ambient magnetic field was imposed and 
shock waves were formed around the interface between the piston plasma and the background plasma.

A shock wave is formed through the gyration of the ambient plasma in the piston plasma region 
in the present study with a weak ambient magnetic field. 
The comparison among the four runs showed that 
a perpendicular collisionless shock 
is formed only when the piston plasma is magnetized. 
It is suggested that 
the spontaneous magnetic field in the piston plasma due to the Biermann battery 
plays an essential role in the formation of a perpendicular collisionless shock 
in the interaction between the two plasmas. 
The shock downstream is formed during the gyration of ambient ions 
in the piston plasma region on the timescale of quarter gyro period of ambient ions.  

\acknowledgments{
This work was supported by JSPS KAKENHI grant numbers 18H01232(RY, TM), 16K17702 (YO), 
15H02154, 17H06202 (YS), 17H18270 (SJT), and 17J03893 (TS).  
This work was also supported partially by the joint research project of the Institute of Laser Engineering, 
Osaka University.
The computer simulations were performed on the CIDAS supercomputer system 
at the Institute for Space-Earth Environmental Research in Nagoya University 
under the joint research program. 
}

\appendix
\section{}
{Figure 5 shows the numerical dispersion relation of the electric field $E_x$ component in Run 1 
obtained by Fourier transformation  for $0.2 < x < 0.3$ cm and $3.9 < t < 4.68$ nsec 
with 12,050 points in position and 3000 points in time. 
The frequency and wavenumber are normalized by the angular cyclotron frequency 
$\omega_{ceb}$ and the inertial length $d_{eb}$ of the background electrons, 
respectively. 
The local electron cyclotron frequency varies 
from $\omega_{ce,local} = \omega_{ceb}$ to $\omega_{ce,local} \approx 35 \ \omega_{ceb}$ 
in both space and time due to the penetration of the Aluminum plasma. 
The spectral enhancement is seen in this frequency range. 
The plasma angular frequency of the background electrons is $\omega_{peb}/\omega_{ceb} \approx 78.74$, 
which is larger than the angular frequency of the excited waves. 
Hence, the waves are not excited at the local upper hybrid resonance frequency 
nor the local electron plasma frequency, but at the local electron cyclotron frequency. 
The phase velocity of the excited wave mode is obtained as 
$v_p \approx 1.5 \ \omega_{ceb}d_{eb} \approx 600$ km/s, which is close to the  
drift velocity of aluminum ions in this region as seen in Fig.1. 
}

\begin{figure}[t]
\center
\includegraphics[width=0.9\textwidth,bb=0 0 60 60]{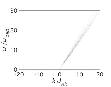}
\caption{
Numerical dispersion relation of the electric field $E_x$ component in Run 1
obtained by Fourier transformation  for $0.2 < x < 0.3$ cm and $3.9 < t < 4.68$ nsec. 
The frequency and wavenumber are normalized by the angular cyclotron frequency 
$\omega_{ceb}$ and the inertial length $d_{eb}$ of the background electrons, 
respectively. 
}
\end{figure}


\begin{thebibliography}{99}

\bibitem{Balogh_2013}
A. Balogh and R. A. Treumann, 
``Physics of Collisionless Shocks,'' (Springer Verlag, New York, 2013).

\bibitem{Burgess_2015}
D. Burgess and  M. Scholer, 
``Collisionless Shocks in Space Plasmas,'' (Cambridge University Press, Cambridge, 2015).

\bibitem{Kuramitsu_2010}
Y. Kuramitsu, Y. Sakawa, T. Morita, C. D. Gregory, J. N. Waugh, S. Dono, H. Aoki, H. Tanji, M. Koenig, N. Woolsey, and H. Takabe, 
Phys. Rev. Lett. \textbf{106}, 175002 (2011).

\bibitem{Morita_2010}
T. Morita,  Y. Sakawa, Y. Kuramitsu, S. Dono, H. Aoki, H. Tanji, T. N. Kato, Y. T. Li, Y. Zhang, X. Liu, J. Y. Zhong, H. Takabe, and J. Zhang,
Phys. Plasmas \textbf{17}, 122702 (2010).

\bibitem{Yuan_2017}
D. Yuan,  Y. Li, M. Liu, J. Zhong, B. Zhu, 
Y. Li, H. Wei, B Han, X Pei, J Zhao, et al.
Sci. Rep., \textbf{7}, 42915 (2017).

\bibitem{Ross_2017}
J. S. Ross, D. P. Higginson, D. Ryutov, F. Fiuza, R. Hatarik, 
C. M. Huntington, D. H. Kalantar, A. Link, B. B. Pollock, B. A. Remington et al.,
Phys. Rev. Lett. \textbf{118}, 185003 (2017).

\bibitem{Schaeffer_2014}
D. B. Schaeffer,  E. T. Everson, A. S. Bondarenko, S. E. Clark, C. G. Constantin, S. Vincena, B. Van Compernolle, S. K. P. Tripathi, D. Winske, W. Gekelman, and C. Niemann,
Phys. Plasmas \textbf{21}, 056312 (2014).

\bibitem{Niemann_2014}
C. Niemann, W. Gekelman, C. G. Constantin, E. T. Everson, D. B. Schaeffer, A. S. Bondarenko, S. E. Clark, D. Winske, S. Vincena, B. Van Compernolle, and P. Pribyl, 
Geophys. Res. Lett. \textbf{41}, 7413 (2014).

\bibitem{Schaeffer_2017a}
D. B. Schaeffer, W. Fox, D. Haberberger, G. Fiksel, A. Bhattacharjee, D. H. Barnak, S. X. Hu, and K. Germaschewski,  
Phys. Rev. Lett. \textbf{119}, 025001 (2017).

\bibitem{Schaeffer_2017b}
D. B. Schaeffer, W. Fox, D. Haberberger, G. Fiksel, A. Bhattacharjee, D. H. Barnak, S. X. Hu, K. Germaschewski, and R. K. Follett, 
Phys. Plasmas \textbf{24}, 122702 (2017).

\bibitem{Shoji_2016}
Y. Shoji, R. Yamazaki, S. Tomita, Y. Kawamura, Y. Ohira, S.Tomiya, Y. Sakawa, T. Sano, Y. Hara, S. Kondo, H. Shimogawara, S. Matsukiyo, T. Morita, K. Tomita, H. Yoneda, K. Nagamine, Y. Kuramitsu, T. Moritaka, N. Ohnishi, T. Umeda, and H. Takabe, 
Plasma Fusion Res. \textbf{11}, 3401031 (2016). 

\bibitem{Fiuza_2012}
F. Fiuza, R. A. Fonseca, J. Tonge, W. B. Mori, and L. O. Silva, 
Phys. Rev. Lett. \textbf{108}, 235004 (2012)


\bibitem{Ruyer_2015}
C. Ruyer, L. Gremillet, and G. Bonnaud,
Phys. Plasmas \textbf{22}, 082107 (2015).


\bibitem{Grassi_2017}
A. Grassi, M. Grech, F. Amiranoff, A. Macchi, and C. Riconda, 
Phys. Rev. E \textbf{96}, 033204 (2017)


\bibitem{Biermann_1950}
L. Biermann, 
Z. Naturforsch. \textbf{5a}, 65 (1950). 

\bibitem{Fiksel_2014}
G. Fiksel, W. Fox, A. Bhattacharjee, D. H. Barnak, P.-Y. Chang, K. Germaschewski, S. X. Hu, and P. M. Nilson,
Phys. Rev. Lett. \textbf{113}, 105003 (2014).


\bibitem{Fox_2018}
W. Fox, J. Matteucci, C. Moissard, D. B. Schaeffer, A. Bhattacharjee, K. Germaschewski, and S. X. Hu, 
Phys. Plasmas \textbf{25}, 102106 (2018).


\bibitem{Stamper_1991}
J. A. Stamper, 
Laser Part. Beams \textbf{9}, 841 (1991).

\bibitem{Gregori_2012}
G. Gregori, A. Ravasio, C. D. Murphy, K. Schaar, A. Baird, A. R. Bell, A. Benuzzi-Mounaix, R. Bingham, C. Constantin, R. P. Drake, M. Edwards, E. T. Everson, C. D. Gregory, Y. Kuramitsu, W. Lau, J. Mithen, C. Niemann, H.-S. Park, B. A. Remington, B. Reville, A. P. L. Robinson, D. D. Ryutov, Y. Sakawa, S. Yang, N. C. Woolsey, M. Koenig and F. Miniati, 
Nature \textbf{481}, 480 (2012).

\bibitem{Kugland_2013}
N. L. Kugland, J. S. Ross, P.-Y. Chang, R. P. Drake, G. Fiksel, D. H. Froula, S. H. Glenzer, G. Gregori, M. Grosskopf, 
G. Huntington et al., 
Phys. Plasmas \textbf{20}, 056313 (2013).

\bibitem{Ryutov_2013}
D. D. Ryutov, N. L. Kugland, M. C. Levy, C. Plechaty, J. S. Ross, and H. S. Park, 
Phys. Plasmas \textbf{20}, 032703 (2013).



\bibitem{Umeda_2006}
T. Umeda and R. Yamazaki, 
%Particle simulation of a perpendicular collisionless shock: 
%A shock-rest-frame model, 
Earth Planets Space \textbf{58}, e41 %--e44. 
(2006). 

\bibitem{Sokolov_2013}
I. V. Sokolov,  
%Alternating-order interpolation in a charge-conserving scheme for
%particle-in-cell simulations, 
Comput. Phys. Commun. \textbf{184}, 320 (2013)

\bibitem{Umeda_2003}
T. Umeda, Y. Omura, T. Tominaga, and H. Matsumoto, 
%A new charge conservation method 
%in electromagnetic particle simulations, 
Comput. Phys. Commun. \textbf{156}, 73 (2003)

\end{thebibliography}
\end{document}